\documentclass[twocolumn,showpacs,preprintnumbers,amsmath,amssymb]{revtex4}

\usepackage{graphicx}
\usepackage{dcolumn}
\usepackage{bm}

\begin{document}


\title{Reverberating activity in a neural network with distributed signal transmission delays}

\author{Takahiro Omi}
\author{Shigeru Shinomoto}%
\affiliation{%
  Department of Physics, Kyoto University, Sakyo-ku, Kyoto 606-8502, Japan  
}%

\date{\today}

\begin{abstract}

It is known that an identical delay in all transmission lines can destabilize macroscopic stationarity of a neural network, causing oscillation or chaos.
We analyze the collective dynamics of a network whose intra-transmission delays are distributed in time.
Here, a neuron is modeled as a discrete-time threshold element that responds in an all-or-nothing manner to a linear sum of signals that arrive after delays assigned to individual transmission lines.
Even though transmission delays are distributed in time, a whole network exhibits a single collective oscillation with a period close to the average transmission delay.
The collective oscillation can not only be a simple alternation of the consecutive firing and resting, but also nontrivially sequenced series of firing and resting, reverberating in a certain period of time.
Moreover, the system dynamics can be made quasiperiodic or chaotic by changing the distribution of delays.
\end{abstract}


\maketitle

\section{Introduction}

A number of model neural networks exhibit collective oscillation.
Their mechanisms of oscillation can be classified into three types:
(1) Individual neurons are oscillators with varied frequencies, and collective oscillation emerges through phase interaction between individual oscillators;
(2) Neurons are simple relaxation elements, and oscillation emerges as a result of interaction between excitatory and inhibitory groups of neurons;
(3) Neurons are instantaneous integrators of incoming signals, and oscillation emerges due to signal transmission delay between elements.
The first two scenarios have been studied in detail for many decades, and are already established as fundamental nonlinear dynamical phenomena \cite{wilson-cowan72,Amari72,Kuramoto84,Sakaguchi87,Gerstner95,Shimokawa06}.
Light was shed on the third scenario rather recently and much is not known mathematically in spite of the ubiquity of signal transmission delays, not only in networks of biological neurons but also in networks of artificial electrocircuits \cite{Choi85,Marcus89,Sano07}.

In biological neural networks, the transmission delay is a sum of axonal, synaptic and dendritic delays.
It has been reported that the delay can be comparable to or longer than somatic membrane time scale \cite{Swadlow85,Pelletier02,Soleng03}.
We consider here a model network in which the transmission delays are distributed in a large interval of time.
We model the neuron for simplicity as a discrete-time threshold element that updates its state at each time step according to the summed input signals, each of which has arrived with a fixed transmission delay.

In the studies of neural networks, synchronous update from the preceding states of other elements has often been discussed \cite{Amari71,Shinomoto86,Kinzel85,Shinomoto87,Kurten88,Greenfield01,Mcguire02,Bornholdt03,Bertschinger04}.
This synchronous update rule can give rise to a period-two macroscopic oscillation as the strength of the inhibitory connections is increased \cite{Amari71,Shinomoto86,Kinzel85,Shinomoto87}.
In the present study, we consider the more general case in which inter-neuronal transmission delays are distributed widely in time.
We derive a recurrence equation of the macroscopic order parameter representing the mean activity from the dynamics of the individual threshold elements.

The network is found to exhibit a collective oscillation with period close to the mean transmission delay.
The collective dynamics can not only be a simple alternation of the consecutive firing and consecutive resting, but also a nontrivially sequenced series of firing and resting, repeating in a given period of time.
For the system whose inter-neuronal transmission delays are distributed uniformly in a given range of time, we obtain multiple stable periodic orbits.
Due to the multi-stability, one can store a nontrivial firing sequence in the network.
For the case of non-uniformly distributed delays, we solve the recurrence equation and find that the network can exhibit not only periodic but also quasiperiodic or chaotic dynamics.

The present paper is organized as follows:
In section II a recurrence equation of the mean activity is derived.
In section III the stationary solution for the macroscopic dynamical equation and its linear stability is examined analytically.
In section IV the macroscopic state equation is solved numerically for a variety of distributions of the delays.
In section V the numerical simulation of the original microscopic equations is carried out and compared with the solution of the macroscopic state equation.
In section VI we discuss the significance of delayed networks.

\section{Derivation of a macroscopic state equation}

\begin{figure}[h]
\begin{center}
\includegraphics[width = 80mm]{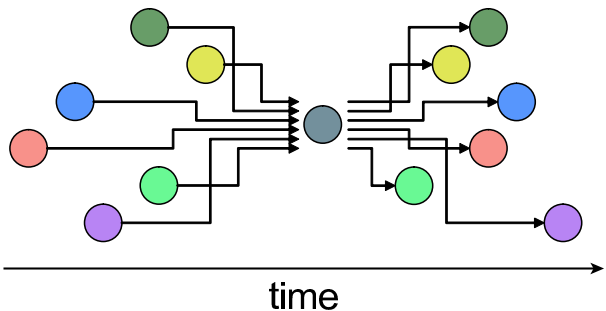}
\caption{
Every neuron receives signals that arrive after delays in the individual transmission lines.
The response signal is sent back to the other neurons with different delays and different connection weights.
}
\label{fig:schematic}
\end{center}
\end{figure}

In this section, we derive the dynamical equation of the macroscopic activity from the microscopic dynamics of the individual elements, each of which is a simple threshold neuron that evokes an all-or-nothing response to an input $v_{i}(t)$ at discrete times $t=0,1,2,3,\cdots$ as,
\begin{equation}
x_i(t)=\textrm{sgn}\left(v_{i}(t)\right),
\label{eq:micro}
\end{equation}
where $\textrm{sgn}(v)$ is the sign function that takes values $+1$, $0$, and $-1$ respectively for $v > 0$, $v=0$, and $v < 0$.
Here, $v_{i}(t)$ is a ``total input'' representing a linear sum of incoming signals from other neurons,
\begin{equation}
v_{i}(t)=\sum_{j=1}^{n}w_{i,j}x_{j}(t-d_{i,j})+s_{i},
\label{eq:v}
\end{equation}
where $w_{i,j}$ is the ``synaptic weight'' that represents excitatory (positive) or inhibitory (negative) connection from the $j$th neuron to the $i$th neuron.
$d_{i,j} (=1,2,3\cdots)$ is the signal transmission delay specified for each transmission line (Fig.\ref{fig:schematic}).
$s_{i}$ will be called the ``external stimulus'' in this paper.
In an asymmetrical representation in which neuronal firing and resting are represented as $(x_{j}+1)/2=1$ or $0$, the external stimulus should be read as $s_{i}-\sum_{j=1}^{n}w_{i,j}$.

A macroscopic order parameter representing the mean activity is defined as the average neuronal state at each time $t$:
\begin{equation}
X(t) \equiv \frac{1}{n}\sum_{i=1}^{n}x_i(t).
\label{eq:activity}
\end{equation}
In the following, we derive a dynamical equation of the macroscopic state, in parallel with Amari's derivation for the synchronous update rule \cite{Amari71}, which corresponds to a particular case of the unit-time delay in the present model, $\{d_{i,j}=1\}$.

The macroscopic state defined above is identical to the difference of ratios of positive and negative total inputs $v_i$ to individual neurons ($i=1,2,\cdots,N$).
Using the distribution $p_t(v)$ of total inputs $v$ at time $t$, the macroscopic state can be represented as
\begin{equation}
X(t)=\int_{0}^{\infty} p_t(v) dv - \int_{-\infty}^{0} p_t(v) dv.
\label{eq:pv}
\end{equation}

The central limit theorem holds for the summed inputs, $\{ \sum_{j=1}^{n} w_{i,j}x_j(t-d_{i,j}) \}_i$, in the limit of a large number of neurons, if the individual signals $\{ w_{i,j}x_j(t-d_{i,j}) \}_{i,j}$ are independently sampled from a given distribution of a finite variance.
In the present deterministic model, the statistical independence holds if individual neuronal states $\{x_{j}\}$ are chosen independently from synaptic connections $\{w_{i,j}\}$.
As the microscopic states are updated through synaptic connections, however, $\{x_{j}\}$ are not independent from $\{w_{i,j}\}$ any more as time goes by.
Even with this condition, there would be room for statistical independence, if the connection weights $\{w_{i,j}\}_{i,j}$ themselves are chosen independently of each other.
The question of statistical independence was initially raised by Rozono\'er \cite{Rozonoer69}.
It was proven by Amari that the statistical independence holds under some special conditions \cite{Amari77}.
In the present paper, we further introduce the distribution of transmission delays $\{d_{i,j}\}$.
This raises another problem of the correlation between $\{x_{j}\}$ and $\{d_{i,j}\}$.
In this paper, we do not go into this open problem, but rather, use the assumption of their statistical independence to construct a macroscopic state equation for our proposed delayed networks.
We will examine the validity of the assumption by comparing the solutions of the macroscopic state equation with the simulation of microscopic equations.

In addition, if the ``external stimuli'' $\{s_{i}\}_i$ are normally distributed, then the distribution $p_t(v)$ of total inputs $v$ is Gaussian, characterized solely by the mean $\mu_t$ and variance $\sigma_t^2$ at each time $t$ as
\[
p_t(v)=\frac{1}{\sqrt{2 \pi \sigma_t^2}} \exp{\left(-\frac{(v-\mu_t)^2}{2 \sigma_t^2}\right)}.
\]
By inserting this into Eq.(\ref{eq:pv}), the macroscopic state equation is obtained as
\begin{eqnarray}
X(t)=F\left(\frac{\mu_t}{\sigma_t}\right),
\label{eq:macro1}
\end{eqnarray}
where $F(x)$ is the error function:
\begin{equation}
F(x)=\sqrt{\frac{2}{\pi}}\int_{0}^{x}e^{-\frac{x^2}{2}}dx.
\end{equation}

Under the above-mentioned assumption of statistical independence, the mean $\mu_t$ and variance $\sigma_t^2$ of input signals $v$ are obtained as
\begin{eqnarray}
\mu_t&=&n\bar{w} \bar{a}(t) + \bar{s}, 
\label{eq:mu} \\
\sigma_t^2&=&n\bar{w}^2 \left(1- \bar{a}(t)^2 \right) + n\sigma^2_w + \sigma^2_s, 
\label{eq:sigma}
\end{eqnarray}
where $\bar{w}$, $\bar{s}$, $\sigma^2_w$ and $\sigma^2_s$ are the means and variances of $\{w_{i,j}\}$ and $\{s_{j}\}$.
$\bar{a}(t)$ denotes the mean past activity:
\begin{equation}
\bar{a}(t) \equiv \frac{1}{n^2} \sum_{i=1}^n \sum_{j=1}^n x_j(t-d_{i,j}).
\label{eq:apast}
\end{equation} 

We consider the case that delays $\{d_{i,j}\}$ are randomly distributed from $d=1$ to $m$ over transmission lines with the ratios $\rho_d \ge 0$, ($\rho_1+\rho_2+\cdots+\rho_m=1$).
Assuming the statistical independence between individual states $\{x_j\}$ and delays $\{d_{i,j}\}$, the mean past activity is given by the weighted average of the past macroscopic states:
\begin{equation}
\bar{a}(t) = \sum_{d=1}^m \rho_d X(t-d).	
\label{eq:meanX}
\end{equation}

The evolution equation of the macroscopic state is given by inserting the mean $\mu_t$ and variance $\sigma_t^2$ into Eq.(\ref{eq:macro1}).
If the model parameters satisfy $n\bar{w}^2 \ll n\sigma_w^2+\sigma_s^2$, the evolution equation simplifies to
\begin{equation}
X(t)=F\left(W\sum_{d=1}^m\rho_dX(t-d)+S\right),
\label{eq:macro}
\end{equation}
where $W$ and $S$ are dimensionless parameters respectively representing the average synaptic weight and the external stimulus,
\begin{eqnarray}
W &=& n\bar{w}/\sqrt{n\sigma_w^2+\sigma_s^2},\label{eq:paraw}\\
S &=& \bar{s}/\sqrt{n\sigma_w^2+\sigma_s^2}. \label{eq:paras}
\end{eqnarray}
We will analyze the recurrence equation (\ref{eq:macro}) in the following sections.
It should be noted that $n\bar{w}^2 \ll n\sigma_w^2+\sigma_s^2$ is not an essential condition for a macroscopic equation (\ref{eq:macro1}) to hold but is merely introduced to make the analysis simpler.

\section{Linear stability analysis of macroscopic stationary states}
Given a macroscopic stationary state $X(t)=X_0$ that satisfies 
\begin{equation}
X_0=F\left(WX_0+S\right),
\end{equation}
we wish to analyze its stability.
For this purpose, the recurrence equation (\ref{eq:macro}) is linearized with respect to the deviation from the stationary state, $\delta X(t) \equiv X(t)-X_0$, as
\begin{equation}
\delta X(t)=\beta\sum_{d=1}^m\rho_d \delta X(t-d),
\label{eq:linear}
\end{equation}
where 
\begin{equation}
\beta=\left.\frac{dF(WX+S)}{dX}\right|_{X=X_0}.
\end{equation}
The stationary state is locally stable if all roots of the characteristic equation,
\begin{equation}
\alpha^m-\beta(\rho_1\alpha^{m-1}+\rho_2\alpha^{m-2}+\cdots+\rho_{m-1}\alpha+\rho_m)=0,
\label{eq:character}
\end{equation}
are smaller than 1 in magnitude.

\subsection{Synchronous update rule, or unit-time delay}

\begin{figure}[h]
\begin{center}
\includegraphics[width = 80mm]{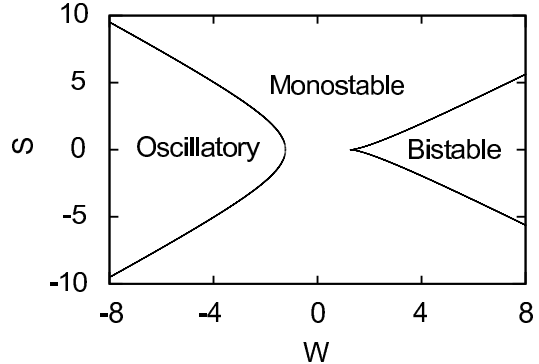}
\caption{
Phase diagram for the synchronous update rule. 
$W$ and $S$ are dimensionless parameters respectively representing the average synaptic weight and external stimulus.
See the text for details.
}
\label{fig:phase-nodelay}
\end{center}
\end{figure}

We start with the synchronous update rule originally studied by Amari~\cite{Amari71}, which corresponds to the case of unit time delay $\{d_{i,j}=1\}$ in the present framework.
The linearized equation for this case is simply given by $X(t)=\beta X(t-1)$.
The stationary solution is stable if $|\beta|<1$, marginal if $|\beta|=1$ and unstable if $|\beta|>1$.
The phase space of parameters $W$-$S$ is divided into three regions according to qualitative differences in the macroscopic state stability (Fig.\ref{fig:phase-nodelay});
[monostable region]: The system has only one stable stationary state ($|\beta|<1$);
[bistable region]: The system has one unstable stationary state with $\beta>1$ and two stable stationary states ($|\beta|<1$);
[oscillatory region]: The system has one unstable stationary state with $\beta<-1$, and one stable oscillatory orbit of period two.

\subsection{Uniformly distributed delays}

Next, we consider the case that delays are uniformly distributed in a given interval of time, $\rho_d=1/m$\ for $d=1,2,\cdots,m$. 
The characteristic equation for this case is
\begin{equation}
\alpha^m-\frac{\beta}{m}\left(\alpha^{m-1}+\alpha^{m-2}+\cdots+\alpha+1\right)=0.\label{eq:character2}
\end{equation}
In the following, we prove that for this particular case the stationary solution is locally stable if $-m<\beta<1$, and unstable otherwise.

\begin{figure}[h]
\begin{center}
\includegraphics[width = 80mm]{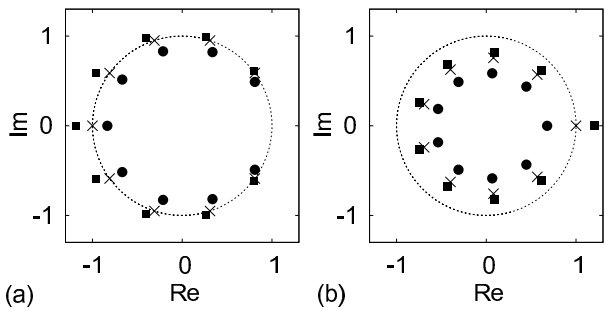}
\caption{
Roots of the characteristic polynomial equation in a complex plane for the case of uniformly distributed delays in an interval of $m=9$.
(a) All the roots leave the unit circle as $\beta$ passes $-m$ from above.
Filled circles, crosses and filled squares represent roots of characteristic equations of stable, marginal and unstable cases, respectively.
(b) One root exceeds $1$ along the real axis, as $\beta$ passes $1$ from below.
}
\label{fig:roots}
\end{center}
\end{figure}
\begin{figure}[h]
\begin{center}
\includegraphics[width = 80mm]{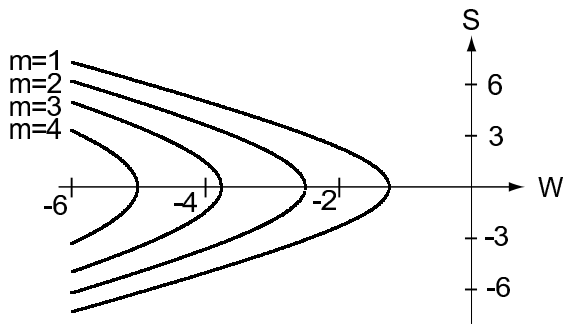}
\caption{
The linear stability boundary between monostable region and oscillatory region shifts to the lower value of $W$ as the distribution interval of the delays $m$ is increased.
}
\label{fig:phase-delay}
\end{center}
\end{figure} 

For $\beta=-m$, all roots of the characteristic polynomial equation (\ref{eq:character2}) align on a unit circle in a complex plane; $\alpha=e^{2\pi ik/(m+1)}\, (k=1,2,\cdots,m)$.
As $\beta$ passes $-m$ from above, all the roots simultaneously leave the unit circle (Fig.\ref{fig:roots}(a)).
For $\beta=1$, the characteristic equation has one root $\alpha=1$, which has the largest length.
As $\beta$ passes $1$ from below, this root exceeds $1$ along the real axis (Fig.\ref{fig:roots}(b)).
In Appendix A, we prove that the characteristic equation (\ref{eq:character2}) possesses no other roots of unit length.
This means that the linearized equation can be destabilized only at these two critical points $\beta = -m$ and $\beta = 1$.

As in the case of the synchronous update rule or unit-time delay $\{ d_{i,j} =1 \}$, the phase space can be divided into three characteristic regions according to qualitative differences in the linear stability.
The distributed transmission delay doesn't shift the boundary between monostable and bistable regions.
The boundary between monostable and oscillatory regions is shifted to the lower direction in $W$ as $m$ is increased (Fig \ref{fig:phase-delay}). 
Note that the present categorization is solely based on the (local) linear stability and there is room for other dynamical orbits to coexist globally.
The coexistence of multiple orbits will be discussed in the next section.

\subsection{Non-uniformly distributed delays}

Finally, we consider the general case that delays are distributed non-uniformly in a given interval of time $m$, satisfying $\rho_1+\rho_2+\cdots+\rho_m=1$.
Even in this general case, the boundary between monostable and bistable regions is the same as the case of uniformly-distributed delays as well as the unit-time delay, as proven below:
At $\beta=0$, all roots of the characteristic equation (\ref{eq:character}) degenerate at $\alpha=0$. 
While for $-1<\beta<1$, the characteristic equation does not have roots of $|\alpha|=1$, implying that all roots are inside the unit circle in the complex plane.
At $\beta=1$, one root arrives at $\alpha=1$, above which a stationary solution becomes unstable.

\begin{figure}[h]
\begin{center}
\includegraphics[width = 80mm]{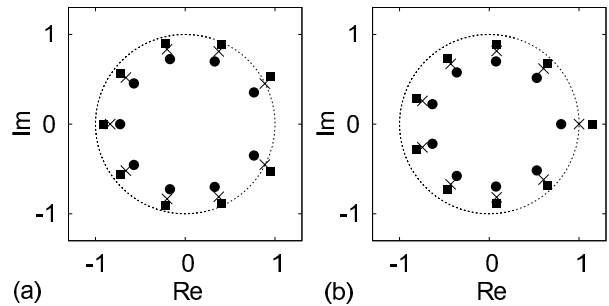}
\caption{
Roots of the characteristic polynomial equation, of an example of nonuniformly distributed delays, $\rho_j=j/45$, $j=1,2,\cdots,9$.
(a) Two complex conjugate roots cross the unit circle while others remain inside, as $\beta$ passes some negative critical value from above.
(b) One root exceeds 1 along the real axis, as $\beta$ passes 1 from below.
}
\label{fig:delay-roots}
\end{center}
\end{figure}

The critical value of $\beta$ on the negative side of $W$ depends on the distribution of the delays $\{\rho_d\}_d$.
As depicted by Fig.\ref{fig:delay-roots}(a), all roots do not necessarily leave the unit circle simultaneously, unlike the uniformly distributed case.

\section{Numerical analysis of the macroscopic state equation}
In this section, we solve the recurrence equation (\ref{eq:macro}) to observe the dynamics of the macroscopic order parameter $X(t)$.
The dynamical state obtained for the synchronous update model or the unit-time delay is a simple period-two oscillation.
Special attention is paid here to nontrivial temporal activity patterns of the network with distributed delays.

\subsection{Uniformly distributed delays}
First, we numerically solve the macroscopic state equation of the network with uniformly distributed delays, $\rho_d=1/m$ for $d=1,2,\cdots, m$:
\begin{equation}
X(t)=F \left(W\frac{\sum_{d=1}^{m}X(t-d)}{m}+S \right). 
\label{eq:macro-uniform}
\end{equation}

\begin{figure}[h]
\begin{center}
\includegraphics[width = 80mm]{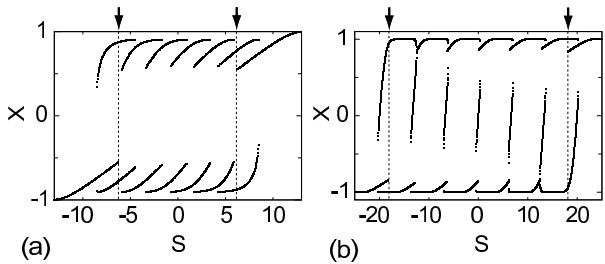}
\caption{
The bifurcation diagrams of the macroscopic recurrence equation (\ref{eq:macro-uniform}) displaying the order parameter $X(t)$ for each value of external stimulus $S$. (a): $W=-10, m=6$; (b) $W=-20, m=6$.
For each value of $S$, we take 100 random initial conditions and plot the last seven $X(t)$.
The linear stability boundaries for (a) and (b) are $S_c=\pm6.2$ and $\pm18.2$, respectively depicted as dashed lines.
}
\label{fig:bifur-diagram}
\end{center}
\end{figure}

Oscillation occurs in the parameter region where inhibition dominates as we see in Figs.\ref{fig:phase-nodelay} and \ref{fig:phase-delay}.
Figure \ref{fig:bifur-diagram} displays the bifurcation diagram of the macroscopic recurrence equation (\ref{eq:macro-uniform}) in the case of $m=6$.
For a given mean connection that is negative, (a) $W=-10$ or (b) $W=-20$, we vary the external stimulus $S$ and observe the temporal activity pattern generated by the recurrence equation:
For each set of parameters of $W$ and $S$, we choose 100 random initial conditions, and iterate the recurrence equation for $t=10000$ and plot the last several values of $X(t)$.

If the external stimulus $S$ is sufficiently small, the system exhibits a stationary sequence of negative $X$ close to $-1$.
As $S$ is increased, a positive $X$ close to $+1$ appears among $X$s close to $-1$, once every $m+1$ iterations.
Note that the oscillation is observed already in the parameter region in which the stationary state is linearly stable.
In other words, oscillatory orbits and a stationary state coexist in the same system, as mentioned in section III.
As $S$ is increased further, a positive $X$ appears twice every $m+1$ iterations.
The temporal order of positive and negative $X$s can be chosen arbitrarily, but it is fixed once the iteration starts.
The proportion of positive $X$s increases with $S$, and eventually the system exhibits a stationary sequence of positive $X$.
Note that different proportions of positive and negative $X$s may coexist for the same value of $S$.

We solved the recurrence equation from initial conditions with $X$s uniformly distributed.
Within our range of numerical investigation, we obtained all possible arrangements of positive and negative $X$s of cycle length $m+1$.
In the case of $m=6$, the period of oscillation is always $m+1=7$, the prime number.
In the general case of $m+1$ that is not a prime number, the period could be shorter with the repetition of short sequenced activity.

Figure \ref{fig:bifur-diagram}(b) shows the solutions for $W=-20$.
It is notable that there appears a new intermediate state of $X$ close to $0$ in addition to states of $X$ close to $+1$ and $-1$.
In comparison with the case $W=-10$, the region in which different proportions of positive and negative $X$s coexist becomes relatively narrower.
In the limit $W\to-\infty$, the proportion of positive and negative $X$s are uniquely determined by the parameter $S' \equiv mS/|W|$, as is proven in Appendix B.

\subsection{Non-uniformly distributed delays}

Next, we examine the case that the delays $\rho_d$ are unevenly distributed:
\begin{equation}
X(t)=F \left(W\frac{\sum_{d=1}^{m}(1+\epsilon_d)X(t-d)}{\sum_{d=1}^{m}(1+\epsilon_d)}+S \right),
\label{eq:macro-non}
\end{equation}
where $\epsilon_d$ is drawn from a Gaussian distribution with mean zero and variance $\epsilon^2$.

\begin{figure}[h]
\begin{center}
\includegraphics[width = 80mm]{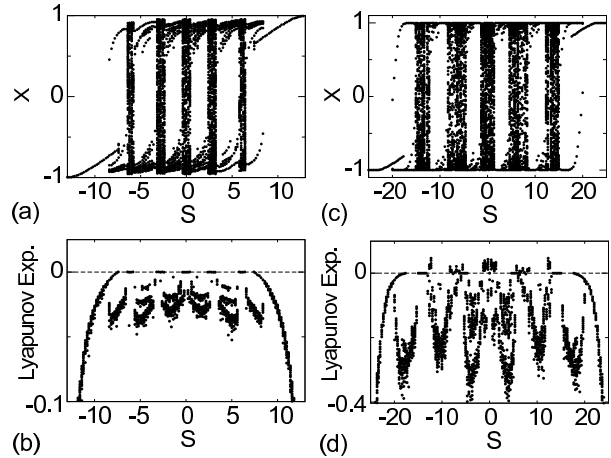}
\caption{Top: The bifurcation diagram of the macroscopic equation (\ref{eq:macro-non}) with bifurcation parameter $S$.
Bottom: the Lyapunov exponents.
(a) and (b): $m=6, W=-10, \epsilon=0.05$; (c) and (d): $m=6, W=-20, \epsilon=0.1$.
For each value of $S$, we take 100 random initial conditions and plot the last seven $X(t)$ for each.
}
\label{fig:bifur-chaos}
\end{center}
\end{figure} 

In addition to periodic oscillation, the system may exhibit quasi-periodic dynamics (Figs.\ref{fig:bifur-chaos}(a) and (b)).
As the deviation $\epsilon$ is increased further, the system may exhibit chaos (Figs.\ref{fig:bifur-chaos}(c) and (d)) characterized by the positive value of the Lyapunov exponent:
\begin{equation}
\lambda=\lim_{t\to\infty} \lim_{\delta \textbf{X}(0)\to 0}\frac{1}{t}\log{\frac{|\delta \textbf{X}(t)|}{|\delta \textbf{X}(0)|}}\,,
\end{equation} 
where $\delta \textbf{X}(t) \equiv (\delta X(t+1), \delta X(t+2),\cdots, \delta X(t+m))$ is an $m$-dimensional perturbation vector added to an original orbit.

\section{Numerical simulation of microscopic dynamics}

In this section, we carry out a numerical simulation of the original microscopic equations (\ref{eq:micro}).
The dynamics of macroscopic order parameter representing the mean activity of microscopic elements are compared with solutions of the macroscopic state equation (\ref{eq:macro}).

\begin{figure}[h]
\begin{center}
\includegraphics[width = 80mm]{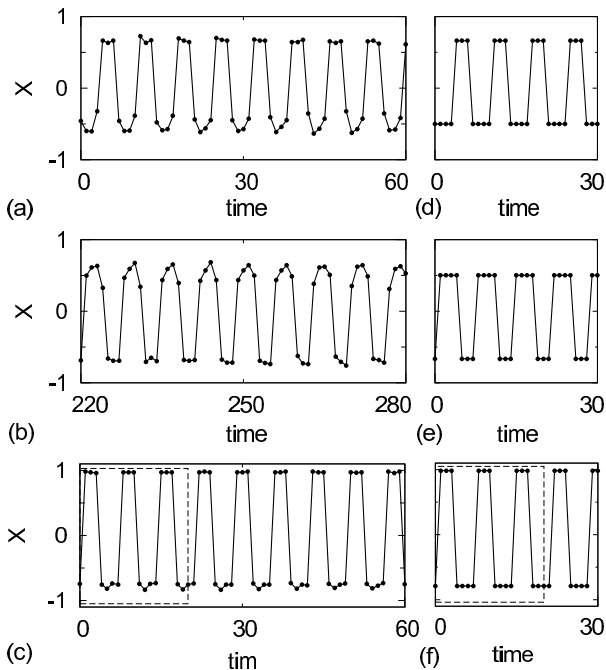}
\caption{Left: The mean activity $X$ obtained by the simulation of the microscopic equation:
(a) and (b): $\bar{w}=-0.08$, $\sigma_w^2=0.09$, $\bar{s}=0$, and $\sigma_s^2=0$;
(c): $\bar{w}=-0.12$, $\sigma_w^2=0.09$, $\bar{s}=0$, and $\sigma_s^2=0$.
Right: Solutions of the macroscopic equation exhibiting similar temporal patterns: (d) and (e): $W=-8.4$ and $S=0$; (f): $W=-12.6$ and $S=0$.
The dashed squares in (c) and (f) are magnified in Figs.\ref{fig:patterns}(a) and (e).
}
\label{fig:micro}
\end{center}
\end{figure} 
\begin{figure}[h]
\begin{center}
\includegraphics[width = 80mm]{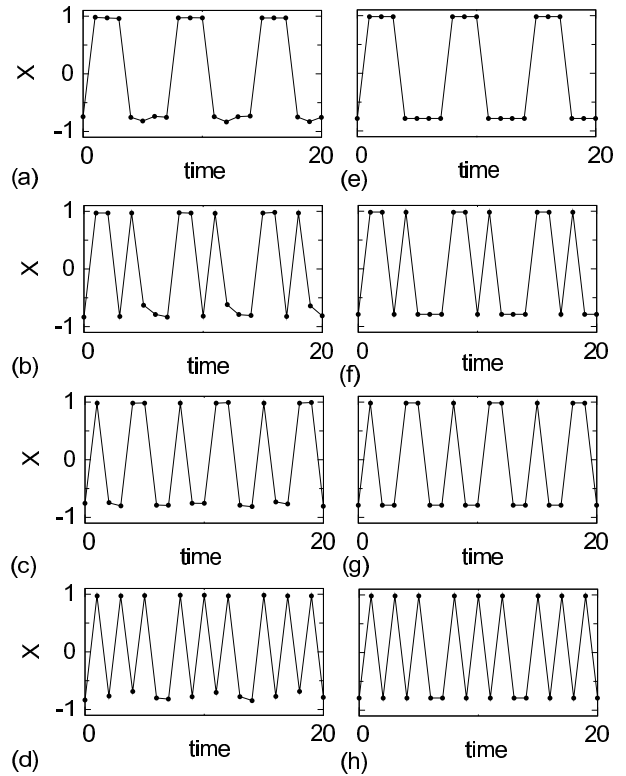}
\caption{Various temporal patterns of mean activity $X$ exhibited by identical networks:
Left: (a) to (d): Numerical simulation of the microscopic equation, with the same parameters of Fig.\ref{fig:micro}(c).
Right: (e) to (h): Solutions of the macroscopic equation, exhibiting temporal activity patterns similar to the simulation results.
 }
\label{fig:patterns}
\end{center}
\end{figure} 

The size of the network we examined is $n=1000$.
Synaptic connections $\{w_{i,j}\}$ are drawn independently from a Gaussian distribution of mean $\bar{w}$ and variance $\sigma_w^2$.
In this simulation, external stimuli $\{s_{i}\}$ are taken as $0$.
The transmission delays $\{d_{i,j}\}$ are chosen randomly from $\{1,2,\cdots,6\}$.
This corresponds to the case of uniformly distributed delays, $\rho_1=\rho_2=\cdots=\rho_6=1/6$

The resulting macroscopic order parameter $X(t)$ is depicted in Fig.\ref{fig:micro}:
(a) and (b) display the mean activity in the case of $\bar{w}=-0.08$, $\sigma_w^2=0.09$, $\bar{s}=0$, and $\sigma_s^2=0$.
The macroscopic state equation (\ref{eq:macro}) with the parameters $W=-8.4$ and $S=0$ estimated with Eqs.(\ref{eq:paraw}) and (\ref{eq:paras}) exhibits temporal mean activity patterns similar to the simulation results (Figs.\ref{fig:micro}(d) and (e)).
It is interesting to observe that the temporal activity pattern is gradually modified as time goes by (from (a) to (b)).
This would be due to the finite size effect.
The oscillation is stabilized in the parameter region of smaller $\bar{w}$, even in a system of the same size.
The simulation result with the parameter $\bar{w}=-0.12$ and a solution of the macroscopic equation with the corresponding parameter $W=-12.6$ are displayed in Figs.\ref{fig:micro}(c) and (f).

As is predicted in the preceding section, the macroscopic order parameter exhibits a wide variety of temporal dynamics depending on the initial condition.
Figs.\ref{fig:patterns}(a)-(d) depict examples of temporal sequences realized in the same network.
Note that these sequenced firing patterns are stable, once they arise from a given initial condition.
Figs.\ref{fig:patterns}(e)-(h) display the similar temporal sequences exhibited by the macroscopic state equation (\ref{eq:macro}).

\section{Discussion}

In the present study, we have demonstrated collective dynamics exhibited by the neural network whose intra-transmission delays are widely distributed in time.
In the case that delays are distributed uniformly in time, the system is found to exhibit collective oscillation with a period close to the average transmission delays.
The network possesses multiple stable orbits of nontrivially sequenced series of firing and resting, reverberating in a certain period of time.
The multistability has also been reported for continuous time dynamical systems that are accompanied with delay \cite{Ikeda87,Losson93,Foss96,Ma07}.
In addition to the periodic motion, we also found that the dynamics can be made quasiperiodic or chaotic by changing the distribution of delays.

The analysis of delayed networks has mainly been confined to their stability \cite{Cao03,Mohamad03,Arik04,Jirsa04,Zhang05}.
Recently, the dynamical aspects of the delayed systems are drawing attention; some possible relevance in biology \cite{Knoblauch03,Izhikevich06,Gong07}; controlling systems by utilizing delays in engineering \cite{Pyragas92,Hikihara96,Hohne07,Kiss07}.
Transmission delays provide networks with potential applications. 
In order to control them, it is necessary to comprehend full aspects of their dynamical characteristics.

\section*{Appendix A}

We prove that the characteristic polynomial equation (\ref{eq:character2})
\[
\alpha^m-\frac{\beta}{m}\left(\alpha^{m-1}+\alpha^{m-2}+\cdots+\alpha+1\right)=0\]
has no roots of unit length, other than the roots $\alpha=e^{2\pi ik/(m+1)}\, (k=1,2,\cdots,m)$ at $\beta=-m$ and $\alpha=1$ at $\beta=1$.

It is readily seen that $\alpha=1$ can be a solution of the characteristic equation, if $\beta=1$.
If $\alpha\ne1$, the characteristic equation can be transformed into
\begin{equation}
\alpha^{m+1}-1=(\frac{\beta}{m}+1)(\alpha^m-1).
\label{eq:character3}
\end{equation}
This equation means that $1$, $\alpha^m$ and $\alpha^{m+1}$ are aligned on a line in the complex plane.
In addition, if $\alpha$ ($\alpha^m$, $\alpha^{m+1}$) lies on the unit circle, at lease two of {$1$, $\alpha^m$ and $\alpha^{m+1}$} must be identical.
This is satisfied with Eq.(\ref{eq:character3}) only if $\alpha^{m+1}=1$ ($\alpha=e^{2\pi ik/(m+1)}, k=1,2,\cdots,m$) and $\beta=-m$.

\section*{Appendix B}

The macroscopic state equation (\ref{eq:macro-uniform}) becomes simpler in the limit of $W\to-\infty$ as
\begin{equation}
X(t)=\textrm{sgn}\left(-\sum_{d=1}^{m}X(t-d)+S'\right),
\label{eq:macro-limit}
\end{equation}
where $S' \equiv mS/|W|$.
We obtain exact solutions of this recurrence equation.
For simplicity's sake, we consider here the case of noninteger $S'$, with which $X(t)$ takes the value of either $+1$ or $-1$.

The recurrence equation generates a stationary sequence of $+1$ if $S' > m$, and a sequence of $-1$ if $S' < -m$.
A sequence composed of both $+1$ and $-1$ is generated if $|S'| < m$.
We prove here that the recurrence equation generates a sequence composed of $+1$s and $-1$s with a period of $m+1$, with the number of $+1$s being $\lceil (m+S')/2 \rceil$:
This sequence of length $m+1$ satisfies the relation
\[\sum_{d=0}^{m}X(t-d)=-m-1+2\lceil (m+S')/2 \rceil.\] 
This can be rewritten as
\begin{equation}
X(t)=-\sum_{d=1}^{m}X(t-d)-m-1+2\lceil (m+S')/2 \rceil.
\end{equation}
For $X=\pm1$, this equation is identical to Eq.(\ref{eq:macro-limit}).
This means that an arbitrary sequence of a period of $m+1$, with the number of $+1$s being $\lceil (m+S')/2 \rceil$ is a solution of the recurrence equation (\ref{eq:macro-limit}).


\section*{Acknowledgments}
We thank Ryota Kobayashi, Shigefumi Hata, Takeaki Shimokawa and Kensuke Arai for stimulating discussions. 
This study is supported in part by Grants-in-Aid for Scientific Research to S.S. from the Ministry of Education, Culture, Sports, Science and Technology of Japan (16300068, 18020015), and by the 21st century COE ``Center for Diversity and Universality in Physics''. 


\end{document}